\pdfoutput=1
\documentclass[reprint,superscriptaddress,preprintnumbers,amsmath,amssymb,aps,prd,tightenlines,longbibliography,nofootinbib]{revtex4-2}

\usepackage{graphicx}
\usepackage[dvipsnames]{xcolor}
\usepackage{amsmath,amssymb,bm,subdepth,mathtools}
\usepackage[colorlinks=true
,urlcolor=blue
,anchorcolor=blue
,citecolor=blue
,filecolor=blue
,linkcolor=red
,menucolor=blue
,linktocpage=true
,pdfa=true
]{hyperref}

\begin{document}

\title{Black Hole Superradiance and the Gravitational Auger Effect}

\author{Joshua~W.~Foster}
\affiliation{Department of Physics, University of Wisconsin-Madison, Madison, WI 53706, USA}

\author{Nicholas~L.~Rodd}
\affiliation{Theory Group, Lawrence Berkeley National Laboratory, Berkeley, CA 94720, USA}
\affiliation{Leinweber Institute for Theoretical Physics, University of California, Berkeley, CA 94720, USA}

\begin{abstract}
Black hole superradiance is a powerful probe of nature's light degrees of freedom.
In the vicinity of a spinning black hole, it exponentially amplifies ultralight bosons from vacuum fluctuations to a macroscopic cloud of enormous occupation number.
Observational superradiance effects -- modified black-hole spins, gravitational-wave and coherent boson emission -- are known to depend sensitively on bosonic self-interactions as induced by a quartic, $\lambda \phi^4$.
Here we point out that purely gravitational interactions within the cloud significantly modify superradiance dynamics, through an analog of the Auger effect.
Moreover, we identify nontrivial evolution of the superradiant levels beyond the first, driven by scattering into even higher levels and their radial overtones.
For the QCD axion, the irreducible gravitational coupling rapidly dissipates the first superradiant level after it reaches saturation, through scalar emission and absorption by the black hole, while truncating the growth of higher levels.
This can dramatically suppress the observability of continuous gravitational-wave signals or signatures in late binary inspirals.
Black hole spin-down remains more robust, but for higher modes begins earlier and proceeds more slowly, requiring revision of existing limits.
\end{abstract}

\maketitle

The existence of a light boson in the spectrum of nature can trigger the superradiant instability of spinning black holes (BHs), causing gravitationally bound scalar clouds to form around them~\cite{Zeldovich:1971, Zeldovich:1972zqp, Press:1972zz, Damour:1976kh, Zouros:1979iw, Detweiler:1980uk, Dolan:2007mj, Arvanitaki:2009fg, Brito:2015oca}.
Superradiance therefore offers a window to light degrees of freedom through observables including BH spin-down, gravitational-wave (GW) production, and modified BH binary inspirals.
In particular, the measured spins of BHs in X-ray binaries and of supermassive BHs~\cite{Arvanitaki:2014wva, Stott:2018opm, Cardoso:2018tly, Davoudiasl:2019nlo, Stott:2020gjj, Unal:2020jiy, Hoof:2024quk, Witte:2024drg}, and the spin distribution of the binary BHs observed by LIGO-Virgo-KAGRA (LVK)~\cite{Ng:2019jsx, Ng:2020ruv, Ning:2026ebu, Aswathi:2025nxa, Khalaf:2026jap}, have been interpreted as providing strong constraints on new states such as the QCD axion.

The superradiance (SR) lifecycle begins when a Kerr BH is born.
A hydrogenic cloud of bosons grows at the expense of the BH mass and spin, continuing until sufficient angular momentum has been extracted to shut off the instability.
The cloud then dissipates through bound-bound annihilation to gravitons, resulting in continuous GW emission~\cite{Arvanitaki:2010sy, Yoshino:2013ofa, Brito:2014wla, Siemonsen:2022yyf}.
Yet the dissipation timescale is long, and competing effects from self-interactions (SIs) can alter the dynamics~\cite{Arvanitaki:2010sy, Yoshino:2012kn, Gruzinov:2016hcq}, as explored systematically for scalars with a quartic $\mathcal{L} \supset \lambda \phi^4/4!$ in Refs.~\cite{Baryakhtar:2020gao, Omiya:2020vji, Omiya:2022mwv, Omiya:2022gwu}.
Reference~\cite{Baryakhtar:2020gao} identified three qualitatively distinct regimes: small, moderate, and large SI.
For small SI, the interactions are irrelevant.
At moderate SI, the initial cloud forms unmodified, but then settles into a quasi-equilibrium with higher levels and dissipates through nonrelativistic scalar emission rather than GWs.
With large SI, the cloud formation saturates early, thereby significantly slowing the BH spin-down.
Sensitivity to new physics therefore depends critically on $\lambda$.

\begin{figure}[!t]
\includegraphics[width=0.9\columnwidth]{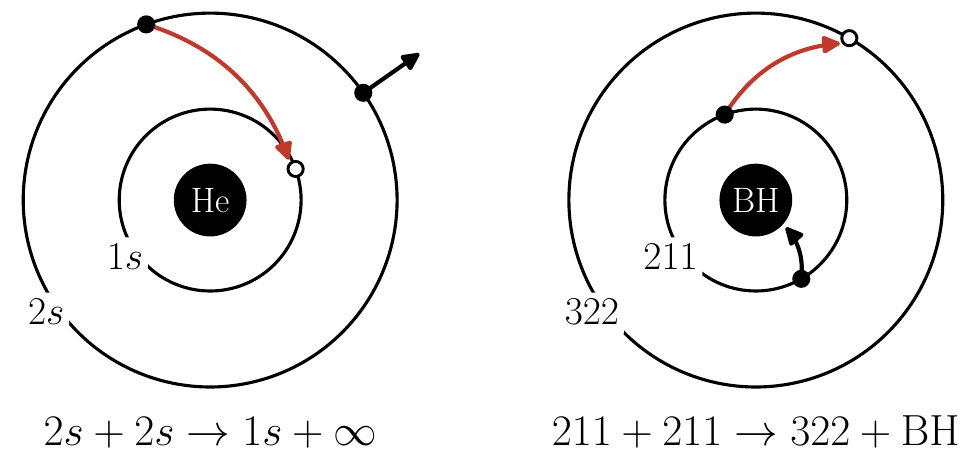}
\vspace{-0.3cm}
\caption{The electromagnetic (left) and gravitational (right) Auger effects.
In helium, the long-range electromagnetic interaction between a pair of $2s$ electrons mediates a transition to a final state with a $1s$ and an unbound electron.
The analog relevant for superradiance is the long-range gravitational interaction between boson cloud states around a BH.
The BH opens a channel where the final-state boson falls into the horizon.
This results in a rapid depletion of the $211$ mode and significantly impacts the superradiant dynamics.}
\vspace{-0.5cm}
\label{fig:auger}
\end{figure}

In this \textit{Letter}, we show there is an effect that dramatically disrupts scalar SR dynamics even at $\lambda=0$: the irreducible self-gravity (SG) of the bosons.
Whilst it is appreciated that SG can shift the hydrogenic energy levels~\cite{Baryakhtar:2020gao, May:2024npn, Cannizzaro:2023jle, Kim:2025wwj, Takahashi:2026ruw, Boskovic:2026eth}, SG also sources interactions amongst bosons similar to a quartic.
SG level transitions are directly analogous to electromagnetically mediated atomic Auger transitions.
We therefore refer to this as the \textit{gravitational Auger effect}; the analogy is highlighted in Fig.~\ref{fig:auger}.

The gravitational Auger effect is strong.
A generic attractive quartic for a scalar with mass $\mu$ and decay constant $f$ is $\lambda = (\mu/f)^2$.
The relevant two-body bosonic potentials are $- G \mu^2/r$ and $- \lambda \delta(\mathbf{r})/8\mu^2$, giving a Fourier-space ratio of $4 \mu^4/\lambda (k M_\mathrm{pl})^2$, where $M_\mathrm{pl} = (8 \pi G)^{-1/2}$.
Evaluating this at the characteristic momentum transfer $k \sim \mu \alpha$ -- $\alpha = G \mu M$ is the gravitational fine-structure constant with $M$ the BH mass -- the two become comparable when $f \sim \alpha M_\mathrm{pl}/2$: a far larger quartic than a naive $f \sim M_\mathrm{pl}$ would suggest.
A full calculation reduces the matching $f$ by a further factor of two; it also reveals that gravity and an attractive quartic \textit{destructively} interfere due to the recoil of the BH in the process.

By extending the two-level SI model studied in Ref.~\cite{Baryakhtar:2020gao} to include SG, we demonstrate that the small SI regime is essentially never realized in scalar SR.
Instead, even with $\lambda=0$, SR proceeds as if in the moderate SI regime.
We extend the analysis to include higher levels and their radial overtones to demonstrate that they cut off the growth of the second superradiant level.

We conclude that the gravitational Auger effect introduces a profound modification to the observational SR prospects for scalars assumed to be in the small SI regime, including the QCD axion.
We demonstrate that SG drives clouds either to dissipate orders of magnitude faster or remain well below saturation, which can remove the expected signatures in late binary inspirals and shorten each continuous GW line, so that $\sim$$10^5$ times fewer lines are expected to be observable at the Advanced LIGO design sensitivity.
Though BH spin-down remains more robust, we find its later stages begin earlier but proceed more slowly, requiring a revision of spin-based constraints that rely on levels beyond $211$.

\vspace{0.2cm}
\noindent {\bf The Auger Effect.}
%
The physics we consider has an ancient atomic analog: the Auger effect~\cite{Meitner:1922, Auger:1923, Auger:1925}.
Consider a charge-$Z$ nucleus surrounded by two electrons in the spin-singlet $\vert 2s,2s \rangle$ state, as shown for helium in Fig.~\ref{fig:auger}.
The effect describes the transition into a $1s$ plus free-electron final state, which we label $X$, mediated by the long-range Coulomb interaction between the electrons.
(The initial state is actually a superposition with $\vert 2p,2p \rangle$; we ignore that complication, although it is an $\mathcal{O}(1)$ correction.)

The calculation of the transition rate is straightforward.
Taking $\alpha_e = e^2/4\pi$ and dimensionless atomic units scaled by the Bohr radius, $1/(Z \alpha_e m_e)$, and the Hartree energy, $(Z \alpha_e)^2 m_e$, the relevant Hamiltonian is
\begin{equation} \label{eq:H_Auger}
H = h_1 + h_2 + \frac{m_e}{m_N} \mathbf{p}_1 \cdot \mathbf{p}_2 + \frac{1}{Z} \frac{1}{|\mathbf{x}_1-\mathbf{x}_2|}.
\end{equation}
Here $1$ and $2$ label the two electrons, $h_i = \mathbf{p}_i^2/2-1/r_i$ are the one-body terms, and what remains is the two-body recoil and Coulomb interaction.
With the nucleus at the origin, the recoil is a vestige of the nucleus one-body term.
Its contribution is negligible.
Given $x \sim p \sim 1$, the recoil is suppressed compared to Coulomb by the ratio of charge-to-mass ratios, $(Z/m_N)/(1/m_e) \ll 1$.
The Auger transition rate is fixed by Fermi's golden rule,
\begin{equation}
\Gamma = 4 \pi \big| \langle 1s,X \vert H \vert 2s,2s \rangle \big|^2 \equiv 4 \pi |\mathcal{M}|^2,
\end{equation}
which includes a factor of two from symmetrizing the final state.
Working in the position basis,
\begin{align} \label{eq:M_Auger}
\mathcal{M} &= \int\! d^3\mathbf{x}\, d^3\mathbf{y}\, \frac{\psi^*_{1s}(\mathbf{x})\, \psi^*_X(\mathbf{y})\, \psi_{2s}(\mathbf{x})\, \psi_{2s}(\mathbf{y})}{Z|\mathbf{x}-\mathbf{y}|} \\
&=\langle X \vert \Phi_{1s2s} \vert 2s \rangle,\quad
\Phi_{1s2s}(\mathbf{y}) \equiv \int d^3\mathbf{x}\,\frac{\psi_{1s}^*(\mathbf{x}) \psi_{2s}(\mathbf{x})}{Z|\mathbf{x}-\mathbf{y}|}. \nonumber
\end{align}
Here $\psi_{1s,2s}$ are hydrogenic wavefunctions, whereas $\psi_X$ is the wavefunction of the outgoing electron as determined by the Coulomb wave equation.
(The outgoing electron is in the $s$-wave and nonrelativistic for $Z \alpha_e \ll 1$.)
The final expression writes the amplitude as an effective one-body transition $2s \to X$ in the presence of the transition density of the second electron.
Defining a source $S(\mathbf{y}) = \Phi_{1s2s}(\mathbf{y}) \psi_{2s}(\mathbf{y})$, the amplitude simplifies further to $\mathcal{M} = \langle X \vert S \rangle$.
However expressed, the integrals evaluate to $\Gamma \simeq 5 \times 10^{-3}\, \alpha_e^2 m_e \sim 0.1\,\mathrm{eV}$, independent of $Z$.

\vspace{0.2cm}
\noindent {\bf Gravitational Transitions.}
%
The equivalent gravitational calculation is conceptually identical.
The wavefunctions are now those of a scalar field in the Kerr BH geometry; we take $\alpha \ll 1$, corresponding to the nonrelativistic, weak-field, Newtonian limit, where the wavefunctions are approximately hydrogenic and labeled $\vert n \ell m \rangle$.
Taking gravitational atomic units -- lengths and energies scaled by $1/\alpha \mu$ and $\alpha^2 \mu$ -- the $N$-scalar Hamiltonian is
\begin{equation} \label{eq:H_G_Auger}
\!H = \sum_i h_i + \sum_{i<j} \!\left[ q \mathbf{p}_i \cdot \mathbf{p}_j - \frac{q}{|\mathbf{x}_i\!-\!\mathbf{x}_j|} - \bar{\lambda} \delta(\mathbf{x}_i\!-\!\mathbf{x}_j) \right]\!.
\end{equation}
This matches Eq.~\eqref{eq:H_Auger} up to the coupling $q = \mu/M$ and the addition of the quartic, with $\bar{\lambda} = \alpha \lambda/8$.
Recoil arises from the BH kinetic term and, for $x \sim p \sim 1$, it is unsuppressed compared to SG; the equivalence principle ensures a unit charge-to-mass ratio.

We consider the transition in Fig.~\ref{fig:auger} between the $211$ and $322$ levels, denoted $1$ and $2$ hereafter.
For initial occupancies $N_1$ and $N_2$, by Fermi's golden rule,
\begin{equation}
\Gamma_H = 2\pi N_1(N_1-1)(N_2+1) \big| \langle X \vert S \rangle \big|^2.
\end{equation}
The prefactors capture the Bose statistics; that $\Gamma_H \neq 0$ even for $N_2=0$ is a consequence of spontaneous scattering.
As for helium, we can express the source in terms of the potential of a scalar transitioning from $1 \to 2$, denoted $V_{21}$.
In the position basis, $S(\mathbf{y}) = V_{21}(\mathbf{y}) \psi_1(\mathbf{y})$, where $V_{21} = q \langle 2 \vert \mathbf{p} \vert 1 \rangle \cdot \mathbf{p} + q \Phi_{21} - \bar{\lambda} \psi_2^* \psi_1$, with $\Phi_{21}$ the transition potential of Eq.~\eqref{eq:M_Auger} at $Z \to -1$.

The final state $X$ falls into the BH, as it must given $E_X = 2E_1 - E_2 = \mu (1-7\alpha^2/36) < \mu$.
$\psi_X$ is again determined from the Coulomb wave equation, with a normalization fixed by the BH horizon, which to leading order in $\alpha$ enters as an absorptive potential $V_\mathrm{abs} = -(i/2) A_H \delta(\mathbf{r})$, with $A_H$ the horizon area.

The results combine to
\begin{equation}\begin{gathered} \label{eq:Gamma_H}
\Gamma_H = \frac{A_H (\alpha \mu)^3}{\pi} N_1(N_1-1)(N_2+1) \big| q \mathcal{I}_G + \bar{\lambda} \mathcal{I}_\lambda \big|^2, \\
\mathcal{I}_G \simeq -0.0914,\quad
\mathcal{I}_\lambda \simeq +0.00184.
\end{gathered}\end{equation}
The opposite signs of the two coefficients are the reason gravity destructively interferes with an attractive quartic.
The origin is a partial cancellation between SG and recoil: the equivalence principle ensures the BH acceleration exactly cancels the uniform part of the SG field, leaving only a tidal component of opposite sign.
(Neglecting recoil, $\mathcal{I}_G \simeq + 0.143$.)
For an attractive quartic $\lambda = (\mu/f)^2$, the scale at which the two cancel is
\begin{equation} \label{eq:f_cross}
f_\times \simeq 0.25\, \alpha M_\mathrm{pl} \simeq 6.1 \times 10^{16}\,\mathrm{GeV}\, (\alpha/0.1).
\end{equation}
Only for $f < f_\times$ does SI dominate over SG.
Whilst $f = f_\times$ tunes this rate to zero, it does not render self-coupling negligible: the ratio $\mathcal{I}_G/\mathcal{I}_\lambda$ is channel-dependent and others, such as $211 + 211 \to 422 + \mathrm{BH}$, do not cancel at $f_\times$ and remain active.

The procedure generalizes to other gravitational Auger transitions.
The rate for $322 + 322 \to 211 + \infty$ is
\begin{equation}\begin{gathered} \label{eq:Gamma_inf}
\Gamma_\infty = 2\pi \alpha^2 \mu N_2(N_2-1)(N_1+1) \big| q \mathcal{I}^{(3)}_G + \bar{\lambda} \mathcal{I}^{(3)}_\lambda \big|^2, \\
\mathcal{I}^{(3)}_G \simeq +0.02995,\quad
\mathcal{I}_\lambda^{(3)} \simeq -3.40 \times 10^{-4},
\end{gathered}\end{equation}
where we have neglected a small correction from the $\ell=5$ partial wave.
Moving forward, we write $\bar{\Gamma}$ for occupancy-stripped rates; for example $\bar{\Gamma}_H \simeq \Gamma_H/N_1^2(N_2+1)$.

\begin{figure}[!t]
\includegraphics[width=\columnwidth]{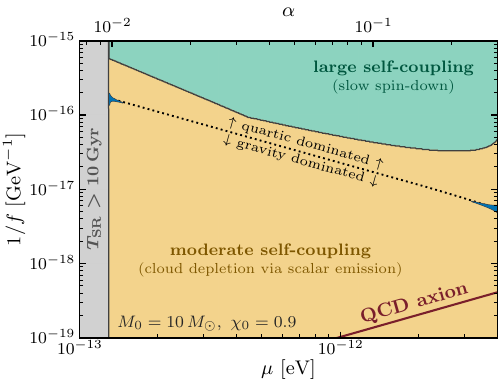}
\vspace{-0.7cm}
\caption{The $211$ self-coupling regimes with SG and an attractive quartic SI, $\lambda = (\mu/f)^2$, for a BH with $M_0 = 10\,M_\odot$ and $\chi_0 = 0.9$ evolved for $T_\mathrm{BH} = 10^{10}\,\mathrm{yr}$.
Even as the quartic vanishes, gravitational Auger transitions place the scalar in the moderate self-coupling regime; small self-coupling survives only in the blue islands at the end of the dotted $f=f_\times$ line.
Above and below that line SI and SG dominate, respectively.
The diagram is almost identical for a repulsive quartic, except the cancellation is absent.
In the gray region there is insufficient time for the $211$ cloud to saturate, and for $\alpha \gtrsim 0.3$, levels beyond $211$ and $322$ become relevant, while the hydrogenic approximations break down~\cite{Baryakhtar:2020gao}.
We also illustrate the $\mu$--$f$ relationship for the QCD axion, which lies entirely within the gravity-dominated moderate self-coupling regime.}
\vspace{-0.5cm}
\label{fig:phase_diagram}
\end{figure}

\vspace{0.2cm}
\noindent {\bf Two-Level Dynamics.}
%
To assess the impact of the gravitational Auger rates, we build on the treatment of Ref.~\cite{Baryakhtar:2020gao}, which studied scalar SR dynamics for $\alpha \ll 1$ in the presence of a quartic SI.
Their analysis focused on the $m=1$ and $2$ modes with the largest SR instabilities, the $211$ and $322$ levels.
Without self-coupling, the two levels grow independently, but a quartic SI allows for scattering processes which transfer pairs of $211$ quanta into $322$ and pairs of $322$ quanta back into $211$, coupling their evolution.
SG drives the same scattering channels and therefore can be studied with an equivalent dynamic system; indeed, for $q = 0$ our Eqs.~\eqref{eq:Gamma_H} and \eqref{eq:Gamma_inf} reproduce the rates of Ref.~\cite{Baryakhtar:2020gao}.
Our results are shown in Fig.~\ref{fig:phase_diagram}.

The occupancies evolve as follows,
\begin{equation}\begin{aligned} \label{eq:rate_equations}
\dot N_1 =\,& \bar{\Gamma}_1 (N_1+1) - 2 \bar{\Gamma}_H N_1^2 (N_2 + 1) \\
&+ \bar{\Gamma}_\infty N_2^2 (N_1 + 1) - 2 \bar{\Gamma}_1^\mathrm{ann} N_1^2, \\ 
\dot N_2 =\,& \bar{\Gamma}_2 (N_2+1) - 2 \bar{\Gamma}_\infty N_2^2 (N_1 + 1) \\
&+ \bar{\Gamma}_H N_1^2 (N_2+1) - 2 \bar{\Gamma}_2^\mathrm{ann} N_2^2.
\end{aligned}\end{equation}
Beyond the $2$-to-$2$ scattering rates controlled by $\bar{\Gamma}_H$ and $\bar{\Gamma}_\infty$, the SR and scalar-to-graviton annihilations scale schematically as~\cite{Detweiler:1980uk,Yoshino:2013ofa},
\begin{equation}\begin{aligned}
\bar{\Gamma}_{n\ell m} &\simeq 2 \mu m^{2\ell+1} C_{n\ell} \alpha^{4\ell+4} \chi \bigg(1-\frac{\mu}{m \Omega_H} \bigg),\\
\bar{\Gamma}^\mathrm{ann}_{n\ell} &\simeq \frac{1}{2} \mu q D_{n\ell} \alpha^{4\ell+9},
\end{aligned}\end{equation}
with $\chi$ the dimensionless spin and $\Omega_H$ the angular velocity of the horizon, $C_{21} = 1/48$ and $C_{32} = 4/885735$, and $D_{21} \simeq 0.011\,(1 - 2.65\,\alpha)$ and $D_{32} \simeq 6.5\times10^{-7}\,(1 - 1.74\,\alpha)$~\cite{Siemonsen:2022yyf} (when moving beyond two levels we take the flat-space coefficients from Ref.~\cite{Yoshino:2013ofa}).
The clouds evolve alongside the BH mass and angular momentum $J = \chi G M^2$,
\begin{align} \label{eq:MJ_evo}
\dot M &= -\mu\big[\bar{\Gamma}_1 (N_1+1) + \bar{\Gamma}_2 (N_2+1)\big] + \mu \bar{\Gamma}_H N_1^2 (N_2 + 1), \nonumber\\
\dot J &= -\big[\bar{\Gamma}_1 (N_1+1) + 2\bar{\Gamma}_2 (N_2+1)\big].
\end{align}
SR extracts energy $\mu$ and angular momentum $m_i$ per quantum (here $1$ or $2$) from the BH, born with mass $M_0$ and spin $\chi_0$, until its spin falls to the energy threshold of the level, $\omega_i = m_i\Omega_H$.
Note that below the SR threshold, $\bar{\Gamma}_i < 0$ and the BH reabsorbs the level, which has no spontaneous counterpart, so that $N_i + 1 \to N_i$ in the SR terms of Eqs.~\eqref{eq:rate_equations} and~\eqref{eq:MJ_evo}.
We now proceed to classify the strength of SG into the self-coupling regimes of Ref.~\cite{Baryakhtar:2020gao}, which are defined by the fate of the $211$ cloud.

We first assume that the $211$ cloud can grow to saturation at $N_1^\mathrm{sat} = \Delta\chi\,GM_0^2$, with $\Delta\chi$ the spin extracted by the mode.
This occurs on a timescale $T_\mathrm{SR} = \bar{\Gamma}_1^{-1} \log (\Delta\chi\,GM_0^2) \simeq 179\, \bar{\Gamma}_1^{-1}$, where the final value holds for $\Delta \chi = 0.5$ and a $10\,M_{\odot}$ BH.
The distinction between small and moderate self-coupling is whether the cloud then dissipates dominantly through GW emission or scattering into $322$.
From Eq.~\eqref{eq:rate_equations}, GW emission acts to deplete the cloud on a timescale $T_\mathrm{GW} = (2\bar{\Gamma}_1^{\rm ann}N_1)^{-1}$.
Meanwhile, scattering feeds $322$ at a rate $\bar{\Gamma}_H N_1^2$ that slows with GW emission, so that the growth continues for $(N_1^\mathrm{sat})^2 \bar{\Gamma}_H T_\mathrm{GW}$ $e$-folds.
Small self-coupling would require this to be fewer than required for $322$ to reach a macroscopic occupation.
As the $322$ cloud grows, the process $322 + 322 \to 211 + \infty$ becomes relevant, until $\Gamma_H$ and $\Gamma_\infty$ balance at $N_2^\mathrm{bal} = \bar{\Gamma}_H N_1^\mathrm{sat}/(2\bar{\Gamma}_\infty)$.
Therefore $322$ requires $\log N_2^{\rm bal}$ $e$-folds to reach balance, so in pure SG with $\Delta \chi=0.5$, the ratio is
\begin{equation} \label{eq:small_margin}
\!\frac{\bar{\Gamma}_H (N_1^\mathrm{sat})^2 T_\mathrm{GW}}{\log N_2^{\rm bal}} \simeq 3.4\times10^5\bigg(\frac{0.1}{\alpha}\bigg)^{7}\bigg(\frac{174}{\log N_2^{\rm bal}}\bigg).
\end{equation}
Accordingly, with gravity alone the system is in the moderate self-coupling regime.
If $\lambda$ is precisely tuned to drive $\bar{\Gamma}_H \to 0$ as in Eq.~\eqref{eq:f_cross}, small self-coupling could re-emerge.
Whether it does depends on the residual size of equivalent channels involving the radial overtones $n22$ with $n \geq 4$, which leaves only the tiny islands in Fig.~\ref{fig:phase_diagram}.
These are the only levels beyond the two-level model that are relevant to classify the phases of $211$.

It remains possible, however, that the $211$ cloud cannot grow to saturation, and SG instead realizes the large self-coupling regime.
Truncated $211$ growth requires two conditions: 1.~$211$ and $322$ reach quasi-equilibrium during the first cloud's growth; and 2.~$211$ is depleted faster than it grows~\cite{Baryakhtar:2020gao}.
For the first condition, by the time $211$ reaches saturation, $322$ grows to an occupancy determined by $\log N_2 = (N_1^\mathrm{sat})^2 \bar{\Gamma}_H/2\bar{\Gamma}_1$.
Sufficient $e$-folds therefore requires $\mathcal{E} \equiv \log N_2 / \log N_2^{\rm bal} > 1$.
The second condition is to ensure $211$ does not continue to grow to saturation, dragging $322$ along with it.
At balance, the rate of depletion of $211$ quanta is $2\bar{\Gamma}_H N_1 N_2^{\rm bal} - \bar{\Gamma}_\infty (N_2^{\rm bal})^2 = \tfrac34(\bar{\Gamma}_H^2/\bar{\Gamma}_\infty)N_1^2$.
Efficient depletion therefore requires $\mathcal{D} \equiv \tfrac34(\bar{\Gamma}_H^2/\bar{\Gamma}_\infty \bar{\Gamma}_1)N_1^2 > 1$, which can be most conservatively evaluated with $N_1 = N_1^\mathrm{sat}$ and $\bar{\Gamma}_1$ evaluated at birth.
We can evaluate the conditions for arbitrary SI as done in Fig.~\ref{fig:phase_diagram}, but with SG alone $\mathcal{E} \propto 1/\alpha$ and $\mathcal{D} \propto \alpha^2$, and we have
\begin{equation}
\mathcal{E}^2\mathcal{D} \simeq \frac{10^2\,\Delta\chi^6}{\chi_0^3(\log N_2^{\rm bal})^2} \lesssim 10^{-3},
\end{equation}
with the BH mass entering only through $\log N_2^{\rm bal}$.
Hence $211$ reaches saturation in the absence of SI, the large self-coupling regime is never realized, and SG always generates a moderately self-coupled SR phenomenology.

As a summary of our findings in Fig.~\ref{fig:phase_diagram}, even without a quartic, gravity places a scalar in the moderate self-coupling regime: the $211$ cloud grows to saturation, but then depletes through self-scattering and scalar emission, rather than decaying through GW emission.
An attractive quartic SI can restore the small self-coupling regime, but only in a narrow range of $f$ around $f_\times$.
For smaller $f$, the quartic dominates and one recovers the regimes of Ref.~\cite{Baryakhtar:2020gao}.
The QCD axion has an attractive quartic $\lambda \simeq 0.35\,(\mu/f_a)^2$ and $\mu \simeq 5.8\,\mu\mathrm{eV}\,(10^{12}\,\mathrm{GeV}/f_a)$~\cite{GrillidiCortona:2015jxo,FlavourLatticeAveragingGroupFLAG:2024oxs}; cancellation requires $\alpha \simeq 0.6\,(M/3\,M_\odot)^{1/2}$, beyond the $\alpha \lesssim 1/2$ at which $211$ is superradiant, so the QCD axion is always in the moderate self-coupling regime.

\begin{figure}[!t]
\includegraphics[width=\columnwidth]{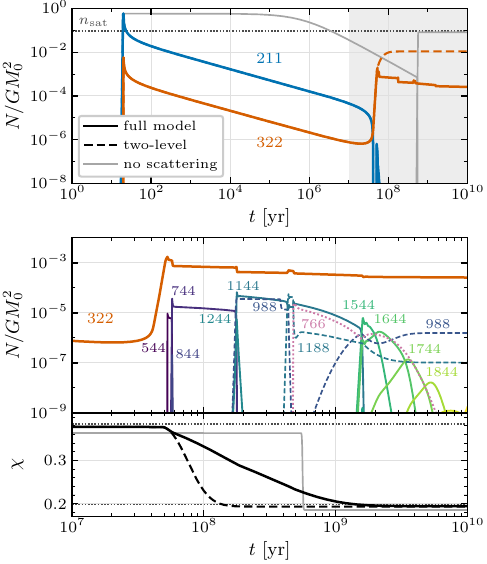}
\vspace{-0.7cm}
\caption{Evolution of a BH born with $M_0 = 10\,M_\odot$ and $\chi_0 = 0.9$, for $\alpha_0 = 0.1$, in the full model (solid), the two-level model (dashed), and without scattering (gray).
\textbf{Top:} occupations of $211$ and $322$, compared with $n_{\rm sat} = N_2^{\rm sat}/GM_0^2$; the shaded window is expanded below.
\textbf{Middle:} $322$ and, in the full model, $544$ and its radial overtones $n44$ (solid), the levels $n88$ (dashed), and $766$ (dotted), colored by $n$.
\textbf{Bottom:} BH spin, with the $211$ and $322$ thresholds at birth (dotted).}
\label{fig:fullmodel}
\vspace{-0.5cm}
\end{figure}

\vspace{0.2cm}
\noindent {\bf Impact of Higher Levels.}
%
While the two-level model allows classification of the superradiant dynamics in the $211$-dominated phase, the later dynamics are richer as scattering amongst higher levels and their radial overtones becomes important.
To study this, we construct a model with thirty-three levels.
Beyond $211$ and $322$, we include: $433$, the next-fastest growing potentially SR level; $544$, $655$, and $766$, which do not grow rapidly by SR but are fed by the scattering of $211$, $322$, and $433$; the radial overtones $644$ through $2044$ of $544$, which we find strongly regulate $322$; and the levels $988$ through $2088$, which can regulate $n44$ and in turn $322$.
We refer to this as a full model, as we demonstrate through convergence tests in the Supplemental Material (SM) that our results for the $211$ and $322$ evolution and the spin-down are robust to the inclusion of further levels.

Focusing on pure SG, we compute rates for all possible scatterings between the levels at leading order in $\alpha$, with resonances regularized as described in the SM, and determine the differential equations that generalize Eqs.~\eqref{eq:rate_equations} and~\eqref{eq:MJ_evo} coupling their evolution.
Otherwise our treatment mirrors that of the two-level model.
For simplicity, we neglect the shifts of the level frequencies by the cloud's self-energy, both in the dynamics and as a drift of the line frequencies.
We also neglect the GW cross-annihilation of pairs of different levels.

An example evolution in the full model is shown in Fig.~\ref{fig:fullmodel}, which again is for SG alone absent any quartic.
SG drains the $211$ cloud to $10^{-3}$ of its peak within $10^5\,\mathrm{yr}$ of saturation, as compared to $5\times10^8\,\mathrm{yr}$ without scattering, while holding $322$ in quasi-equilibrium with it.
Eventually $211$ is sufficiently drained that it exits the balanced regime.
$322$ then grows and, through the absorption channel, removes the remaining $211$.
It also populates the $n44$ levels.
The overtones take turns reaching quasi-equilibrium with $322$, holding it well below its saturated occupation and leading to a considerably slower spin-down to the $322$ threshold.

\vspace{0.2cm}
\noindent {\bf Observational Implications.}
%
Our results reveal small self-coupling is a tuned scenario.
Instead, for $f > f_\times$ -- a region that includes the QCD axion -- SR proceeds through SG-driven moderate self-coupling.
Here we briefly assess the observational implications of this statement, focusing again on the case of pure SG.

\begin{figure}[!t]
\includegraphics[width=\columnwidth]{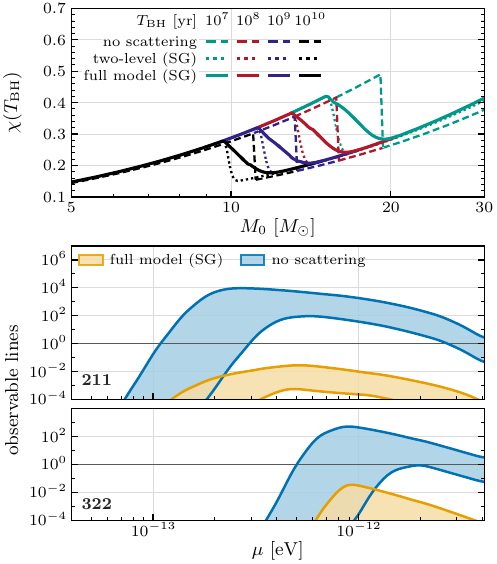}
\vspace{-0.7cm}
\caption{Observables for a scalar with negligible quartic SI, such as the QCD axion, without scattering or with SG.
\textbf{Top:} Regge trajectories for BHs born with $\chi_0 = 0.99$ and variable mass in the presence of a $\mu = 10^{-12}\,\mathrm{eV}$ scalar.
The drop marks the lightest BH that $322$ spins down within $T_\mathrm{BH}$, and occurs gradually in the full model.
At high mass SG leaves the spin slightly higher, as absorbed partners return two-thirds of the SR cloud energy to the BH, raising its threshold spin.
\textbf{Bottom:} expected number of observable $211$ and $322$ GW annihilation lines, without scattering (blue) or with SG in the full model (orange).
Each band spans the BH population models of Ref.~\cite{Baryakhtar:2020gao}, and counts BHs born with $\alpha_0 \le 0.3$ (beyond this our hydrogenic rates are likely unreliable).}
\label{fig:observables}
\vspace{-0.5cm}
\end{figure}

Consider first BH spin-down.
The $211$ cloud first grows to saturation, so that the initial spin-down matches the small self-coupling expectation up to a small shift in the spin-down floor due to the BH's mass growth via scalar absorption~\cite{Baryakhtar:2020gao}.
The subsequent spin-down, driven by $322$, varies considerably between the two-level and full models.
As illustrated in Fig.~\ref{fig:fullmodel}, $322$ has a head start in its SR growth compared to the case without interactions because it is populated by the gravitational Auger scatterings of $211$.
However, the $n44$ overtones hold it far below its saturated occupation, at $\lesssim 10^{-4}$ of the BH mass, in effect placing $322$ in the large self-coupling regime.
As a result, $322$-driven spin-down begins earlier, but proceeds slower.
The top panel of Fig.~\ref{fig:observables} illustrates both these effects through the Regge trajectories~\cite{Arvanitaki:2010sy, Arvanitaki:2014wva}, the spin to which SR reduces a BH of a given mass within a time $T_\mathrm{BH}$ after formation.

The two effects push spin-based constraints in opposite directions, so their net impact is unclear, and quantifying it would require a full population-level analysis of BH spins comparable to Refs.~\cite{Arvanitaki:2016qwi, Ng:2019jsx, Fernandez:2019qbj, Stott:2020gjj, Ng:2020ruv, Cheng:2022jsw, Ning:2026ebu}.
SG leaves the spin evolution of BHs heavier than $\sim$$10^3\,M_\odot$ unchanged for $\alpha \lesssim 0.1$, as the $322$ stage sets in only after a Hubble time for such masses.
Constraints on lighter scalars from supermassive BHs that depend only on spin-down by $211$~\cite{Davoudiasl:2019nlo, Unal:2020jiy, Hoof:2024quk} are therefore largely unaffected.
Those that invoke $322$ and higher levels~\cite{Cardoso:2018tly, Stott:2018opm, Stott:2020gjj} could be modified at the upper end of their mass reach.

SR signatures in late binary inspirals and from continuous GW emission are modified more dramatically, since both depend on a long-lived and abundant cloud.
SG alone allows for a rapid depletion of the cloud (see Fig.~\ref{fig:fullmodel}), which we can estimate in the two-level model.
Once $322$ reaches quasi-equilibrium with $211$, two-thirds of the energy leaving $211$ is absorbed by the BH and one-third escapes as nonrelativistic scalars.
The quasi-equilibrium depletion, $\dot N_1 = -\tfrac34(\bar{\Gamma}_H^2/\bar{\Gamma}_\infty)N_1^3$, drains $211$ from its saturated occupation $N_1^{\rm sat} = \Delta\chi\,GM_0^2$ as $N_1 = N_1^{\rm sat}(1 + 2t/T_\mathrm{SG})^{-1/2}$.
The depletion time $T_\mathrm{SG} = 4\bar\Gamma_\infty/3\bar\Gamma_H^2(N_1^{\rm sat})^2$ and the annihilation time $T_\mathrm{GW}$ scale as follows for $\alpha \ll 1$,
\begin{equation}\begin{aligned} \label{eq:lifetimes}
T_\mathrm{GW} &\simeq 4\times10^5\,\mathrm{yr} \,\bigg(\frac{M}{10\,M_\odot}\bigg) \bigg(\frac{0.5}{\Delta\chi}\bigg) \bigg(\frac{0.1}{\alpha}\bigg)^{15}, \\
T_\mathrm{SG} &\simeq 0.3\,\mathrm{yr}\, \bigg(\frac{M}{10\,M_\odot}\bigg) \bigg(\frac{0.5}{\Delta\chi}\bigg)^2 \bigg(\frac{0.1}{\alpha}\bigg)^{11},
\end{aligned}\end{equation}
so that $T_\mathrm{SG} \ll T_\mathrm{GW}$ and the dominant emission is nonrelativistic scalars, the signal of the BH scalar sirens of Ref.~\cite{Gavilan-Martin:2026zzw}, though generated now by SG rather than SI.
Longer lifetimes can be achieved for smaller $\alpha$, but the GW emission power at saturation decreases rapidly as $P_\mathrm{GW} \propto \alpha^{16}$.
As a result, $211$ clouds are either long-lived but small and weak in GW emission or large and intense in GW emission but short-lived.
The full model reproduces the efficient $211$ depletion (see Fig.~\ref{fig:fullmodel}), leaving a long-lived but underabundant $322$ cloud.

The implications of SG for stellar-mass BHs are most severe.
A cloud that is long-lived compared to typical stellar-mass binary delay times of order $10\,$Myr to $10\,$Gyr~\cite{LIGOScientific:2016vpg, Fishbach:2021mhp}, which requires $\alpha \lesssim 0.02$, holds at most $\sim$$1\%$ of the BH mass, limiting its late-inspiral signatures~\cite{Baumann:2018vus, Baumann:2019ztm, Baumann:2022pkl, Tomaselli:2024faa, Zhang:2018kib, Hannuksela:2018izj, Tomaselli:2023ysb, Tomaselli:2024bdd, Baumann:2021fkf, Tomaselli:2024dbw, Boskovic:2024fga, Boskovic:2025ixx, Tong:2022bbl, Fan:2023jjj, Zhang:2019eid, Brito:2023pyl, Xu:2026cky}.
Larger clouds would have to form shortly before the merger, when their superradiant build-up is likely to be disrupted or interrupted by the binary companion itself~\cite{Berti:2019wnn, Baumann:2019ztm, Cardoso:2020hca, Tomaselli:2024bdd}.
The underabundant $322$ cloud likewise yields only weak inspiral signatures.

The detectability of the GW annihilation $211$ and $322$ lines from stellar-mass BHs is similarly suppressed, as $N_1$ has a stronger depletion channel and $N_2$ does not reach saturation.
We quantify this with our full model in the bottom of Fig.~\ref{fig:observables}, which shows the expected numbers of $211$ and $322$ continuous GW lines observable at $5\sigma$ in an all-sky search at Advanced LIGO design sensitivity, following the population model and search procedure of Ref.~\cite{Baryakhtar:2020gao}, with the Galactic BH distribution of Ref.~\cite{Zhu:2020tht}, and counting BHs born with $\alpha_0 \le 0.3$ (cf.~Ref.~\cite{Ellis:2026gwt}).
Without scattering, up to $10^4$ $211$ lines are expected to be detectable, and at least one for $3\times10^{-13} \lesssim \mu \lesssim 3\times10^{-12}\,\mathrm{eV}$ even under the least favorable population assumptions, whereas with SG fewer than $0.05$ are expected at any $\mu$.
Similarly, fewer than $0.1$ lines are expected to be detectable from $322$ for the most optimistic populations, compared to $\sim$$10^3$ without scattering, almost entirely from BHs born with $\alpha_0 > 0.2$, where our hydrogenic rates are least reliable.
Restricting to $\alpha_0 \le 0.2$ leaves the maximum number of observable $211$ lines unchanged, but reduces that of $322$ lines with SG to $\sim$$10^{-4}$.
Although existing LVK searches may have overstated their reach~\cite{Isi:2018pzk, Sun:2019mqb, Palomba:2019vxe, LIGOScientific:2021rnv}, next-generation detectors such as the Einstein Telescope~\cite{Punturo:2010zz, Hild:2010id, ET:2019dnz, ET:2025xjr} and Cosmic Explorer~\cite{Reitze:2019iox, Evans:2023euw, Ellis:2026gwt} could restore partial sensitivity~\cite{Collaviti:2024mvh}.
The $211$ signature is short lived, and therefore depends sensitively on the recent BH birth rate, whereas the $322$ emission is longer lived but weaker, and its observability turns on GW sensitivity.

For more massive BHs, the relevant timescales grow.
The corresponding increase in the window where a macroscopic $211$ cloud is present could boost the observational prospects compared with stellar-mass BHs.
Higher masses are not without challenges: for example, whether the superradiance growth can occur within a Hubble time, and whether interaction with the local stellar environment around supermassive BHs can reduce the cloud lifetimes~\cite{Cheng:2026dsq}.
Where the balance lies remains to be determined.

\vspace{0.2cm}
\noindent {\bf Discussion.}
%
This \textit{Letter} demonstrates that gravitational interactions of scalar clouds, in close analogy to the atomic Auger effect, can significantly alter SR and its observational consequences.
While SR phenomenology has been most thoroughly explored in the small self-coupling limit, unless the strength and sign of a scalar quartic are fine-tuned to high precision to cancel SG effects, self-coupling is never negligible.
As a result, projected sensitivities and observational constraints which have been developed under the assumption of negligible self-coupling must be revised, including for the QCD axion.
SG also causes the dominant scalar clouds to either dissipate orders of magnitude faster than previously expected or remain far below saturation, such that current searches for continuous GW signals or modified BH binary inspirals can lose virtually all expected constraining power, while its net effect on spin-based constraints remains to be quantified.

In addition to refined observational implications, it will be important to understand how our conclusions extend beyond our perturbative treatment that assumed the system is nonrelativistic, with hydrogenic levels and Newtonian SG.
Larger scalar masses and the associated larger-$\alpha$ regime demand a full Kerr treatment, with the cloud's self-gravitational interaction included either through a sufficiently high-order post-Newtonian calculation~\cite{Namjoo:2017nia, Salehian:2021khb} or with relativistic perturbation theory~\cite{Cannizzaro:2023jle}.

\medskip
\noindent {\it Acknowledgments.}
%
We thank J.~Benabou, D.~Blas, S.~Ellis, and H.~Witek for helpful comments.
ChatGPT and Claude were used throughout the production of this work.
The authors validated and retain full responsibility for all results.
The research of NLR was supported by the Office of High Energy Physics of the U.S.~Department of Energy under contract DE-AC02-05CH11231.
The research of JWF was supported by the U.S.~Department of Energy under award DE-SC0017647.
This work made use of computing resources provided by the Center for High Throughput Computing at the University of Wisconsin, Madison~\cite{https://doi.org/10.21231/gnt1-hw21}, resources of the National Energy Research Scientific Computing Center (NERSC), a Department of Energy User Facility using NERSC award HEP-ERCAP 0037471, and the Lawrencium computational cluster resource provided by the IT Division at the Lawrence Berkeley National Laboratory (supported by the Director, Office of Science, Office of Basic Energy Sciences, of the U.S.~Department of Energy under Contract No.~DE-AC02-05CH11231).

\bibliographystyle{utphys}
\bibliography{refs}

\clearpage
\onecolumngrid

\begin{center}
\textbf{\large Black Hole Superradiance and the Gravitational Auger Effect} \\ 
\vspace{0.05in}
{ \it \large Supplemental Material}\\ 
\vspace{0.05in}
{}
{Joshua~W.~Foster and Nicholas~L.~Rodd}
\end{center}

\twocolumngrid

\setcounter{equation}{0}
\setcounter{figure}{0}
\setcounter{table}{0}
\setcounter{section}{0}
\setcounter{page}{1}
\setcounter{secnumdepth}{2}
\renewcommand{\theequation}{S\arabic{equation}}
\renewcommand{\thefigure}{S\arabic{figure}}
\renewcommand{\thetable}{S\arabic{table}}
\onecolumngrid

This Supplemental Material provides additional details on the calculation of the self-gravitational (SG) scattering rates of the full model discussed in the main text.
Namely, we generalize the two-level calculation to arbitrary pairs of levels and partner partial waves and account for the resonances that are generic given the degeneracy of the hydrogenic spectrum we adopt.
We largely follow and extend the Green's-function treatment of an oscillating source in a gravitational atom of Ref.~\cite{Takahashi:2024fyq}.
Starting in Sec.~\ref{sm:sec:setup} we establish the formalism to determine the SG scattering channels, and then explicitly evaluate the emission and absorption rates.
In Sec.~\ref{sm:sec:resonances} we treat resonances; we reduce the rates to those of the two-level model of the main text in Sec.~\ref{sm:sec:twolevel}, assemble the required rate equations in Sec.~\ref{sm:sec:rates}, and describe our convergence tests for which levels to include in the full model in Sec.~\ref{sm:sec:screen}.

As in the main text, our conventions are $\hbar = c = 1$, $\alpha = GM\mu$, and $\bar\Gamma$ corresponds to occupancy-stripped rates.
Throughout we use dimensionless Bohr units $x = r/a_0$, with the gravitational Bohr radius $a_0 = 1/(\mu\alpha)$.
Energies in units of $\mu\alpha^2$ are written in lowercase, so the level $n$ has binding energy $E_n = \mu\alpha^2 e_n$ with $e_n = -1/(2n^2)$.

\section{Self-Gravity and Recoil in a Multi-Level Cloud}
\label{sm:sec:setup}

In the main text we described the calculation of the rate for $211 + 211 \to 322 + X$.
Here we detail the formalism required to extend this to the complete set of transitions included in our full multi-level model.
We start with a real scalar $\phi$ of mass $\mu$ and focus on the self-coupling induced by gravity.
For simplicity of presentation, we neglect a quartic coupling in our discussion, although its inclusion is a straightforward extension of the logic.
At leading order in $\alpha$ and first order in the mass fraction of the cloud, gravity enters only through the Newtonian gravitational potential $\Phi$ of the BH \textit{and scalar cloud}.
The Klein--Gordon equation then reduces to
\begin{equation}
\ddot\phi - \nabla^2\phi + \mu^2\big(1 + 2\Phi\big)\phi = 0 .
\end{equation}
Here $\phi$ is treated as a classical field, in contrast to the many-body Hamiltonian and Fermi's golden rule approach of the main text.
Although the two descriptions need not agree in general (see e.g.~Refs.~\cite{Bao:2025nsd,Bao:2026prd}), they do for the rates computed here, up to the spontaneous $+1$ in the Bose factors, which we restore in Sec.~\ref{sm:sec:rates}.

The bound levels oscillate at frequencies within $\mu\alpha^2$ of the rest mass, so we factor out the fast oscillations
\begin{equation}
\phi = \frac{1}{\sqrt{2\mu}}\big(\Psi\,e^{-i\mu t} + \Psi^*e^{i\mu t}\big),
\end{equation}
thereby defining a complex field $\Psi$ that varies on the much longer timescale $1/(\mu\alpha^2)$.
As defined, $|\Psi|^2$ is the number density of scalar quanta, and $\mu|\Psi|^2$ is the associated energy density.
In the nonrelativistic limit, the complex field obeys the Schr\"odinger equation
\begin{equation}\begin{gathered} \label{eq:schrodinger}
i\partial_t\Psi = \Big[-\frac{\nabla^2}{2\mu} + \mu\Phi\Big]\Psi , \\
\Phi(\mathbf r, t) = -\frac{GM}{|\mathbf r - \mathbf X(t)|} + \Phi_\mathrm{cl}(\mathbf r, t), \qquad
\nabla^2\Phi_\mathrm{cl} = 4\pi G\mu\,|\Psi|^2 ,
\end{gathered}\end{equation}
where $\mathbf X$ is the position of the BH and $\Phi_\mathrm{cl}$ is the potential sourced by the cloud.

The BH is not held fixed but falls in the potential of the cloud,
\begin{equation} \label{eq:BH_EoM}
\ddot{\mathbf X}(t) = -\nabla\Phi_\mathrm{cl}(\mathbf X, t),
\end{equation}
which together with Eq.~\eqref{eq:schrodinger} conserves the total momentum of the BH and cloud.
A convenient way to account for the motion of the BH is to move to the accelerating frame where the BH is pinned to the origin.
This requires a time-dependent translation and boost of the field $\Psi$,
\begin{equation}
\Psi(\mathbf X + \mathbf r, t) = \exp\Big[i\mu\dot{\mathbf X}\cdot\mathbf r + \frac{i\mu}{2}\int^t dt'\,\dot{\mathbf X}^2\Big]\,\psi(\mathbf r, t),
\end{equation}
where $\mathbf r$ is now the position relative to the BH.
The transformation gives each quantum the momentum $\mu\dot{\mathbf X}$ of the BH and leaves the density $|\psi|^2 = |\Psi|^2$ unchanged.
Inserting the transformation into Eq.~\eqref{eq:schrodinger}, the Schr\"odinger equation for $\psi$ is near identical, but with the BH now at the origin and the appearance of an additional inertial potential $\mu\ddot{\mathbf X}\cdot\mathbf r$, felt by the scalar field in the accelerating frame of the BH, which is the recoil.
(The choice of frame matches that used in Ref.~\cite{Kim:2025wwj}.)
Using Eq.~\eqref{eq:BH_EoM}, the field obeys
\begin{equation}\begin{gathered} \label{eq:eom_nr}
i\partial_t\psi = \Big[H_0 + \mu\Phi_\mathrm{cl}(\mathbf r) - \mu\,\mathbf r\cdot\nabla\Phi_\mathrm{cl}(0)\Big]\psi , \\
H_0 = -\frac{\nabla^2}{2\mu} - \frac{GM\mu}{r}, \qquad
\nabla^2\Phi_\mathrm{cl} = 4\pi G\mu\,|\psi|^2 .
\end{gathered}\end{equation}
Throughout the SM we work in the frame where the BH is fixed at the origin.
This is the frame of standard calculations of scalar cloud levels, their superradiant growth, their annihilation to gravitons, and their scattering through a quartic~\cite{Detweiler:1980uk, Baumann:2019eav, Yoshino:2013ofa, Baryakhtar:2020gao, Siemonsen:2022yyf}.
Those works treat the BH as infinitely massive with the horizon at the origin, and agree with our results to zeroth order in $\mu/M$; the recoil and the cloud's gravity enter at first order.
Since the BH falls freely in the field of the cloud, the recoil contribution removes the uniform part of the field exactly, so that the potential the scalar experiences is $\Phi_\mathrm{cl}(\mathbf r) - \mathbf r\cdot\nabla\Phi_\mathrm{cl}(0)$.
This consequence of the equivalence principle implies that only the tidal part of the potential remains, and is why the gravitational and attractive quartic contributions in Eq.~\eqref{eq:Gamma_H} have opposite signs.
Note that in the main text we work in the inertial center-of-momentum frame; in particular in Eq.~\eqref{eq:H_G_Auger} the form of the recoil contribution, $\mathbf{p}_i \cdot \mathbf{p}_j$, arises from the BH kinetic term.

In our full model, the cloud is a sum over all levels included, which we label $a = (n_a, \ell_a, m_a)$.
Therefore, the field decomposes as
\begin{equation} \label{eq:levels}
\psi = \sum_a c_a\, f_a(\mathbf r)\, e^{-iE_at}, \qquad
f_a = a_0^{-3/2}\,F_a(\mathbf x), \qquad F_a = R_{n_a\ell_a}(x)\,Y_{\ell_am_a}(\hat{\mathbf x}),
\end{equation}
with $R_{n\ell}$ and $Y_{\ell m}$ the hydrogenic radial functions and spherical harmonics, respectively.
The functions are orthonormal, $\int d^3\mathbf{r}\,f_a f_b^* = \delta_{ab}$, leaving the occupancies to be controlled by the coefficients, $|c_a|^2 = N_a$.
The density is then a double sum over pairs of levels,
\begin{equation}
|\psi|^2 = \sum_{a,b} c_a^*\,c_b\,f_a^*f_b\,e^{-i(E_b - E_a)t} .
\end{equation}
Poisson's equation is linear, so the potential is the same double sum, with each term sourced by one product $f_a^*f_b$.
Using $\nabla_r^2 = a_0^{-2}\nabla_x^2$ and $f_a^*f_b = a_0^{-3}F_a^*F_b$,
\begin{equation} \label{eq:Phic}
\Phi_\mathrm{cl}(\mathbf r, t) = \frac{G\mu}{a_0}\sum_{a,b} c_a^*\,c_b\,e^{-i(E_b - E_a)t}\,\Phi_{ab}(\mathbf x),
\qquad \nabla_x^2\Phi_{ab} = 4\pi F_a^*F_b ,
\end{equation}
with $\Phi_{ab}$ dimensionless; it is identical to $\Phi_{21}$ used in the gravitational Auger calculation in the main text (cf.~Eq.~\eqref{eq:M_Auger}).
The diagonal terms $a = b$ are the static potentials of the levels, and the terms with $a \neq b$ are the beats between pairs of levels, which oscillate at the difference of their frequencies.
The recoil contribution of Eq.~\eqref{eq:eom_nr} can be included straightforwardly: the potential in each term is replaced by $\hat\Phi_{ab} = \Phi_{ab} - \mathbf x\cdot\nabla\Phi_{ab}(0)$.

\subsection{Scattering Channels and Sources}

We can now determine what scattering channels exist between the various cloud levels.
Combining Eqs.~\eqref{eq:eom_nr}, \eqref{eq:levels}, and~\eqref{eq:Phic} results in the following triple sum over levels,
\begin{equation} \label{eq:cubic}
\mu\big[\Phi_\mathrm{cl} - \mathbf r\cdot\nabla\Phi_\mathrm{cl}(0)\big]\psi
= \frac{G\mu^3\alpha}{a_0^{3/2}}\sum_{b,c,d} c_c\,c_d\,c_b^*\,\hat\Phi_{bc}(\mathbf x)\,F_d(\mathbf x)\,
e^{-i(E_c + E_d - E_b)t}.
\end{equation}
Not all expressions in the sum enter with a unique coefficient combination and oscillation frequency, and we group those that align.
All terms with $c=d$ are unique, but when $c \neq d$ there are two terms to combine; as an explicit example, $(b,c,d) = (433,211,322)$ and $(433,322,211)$ have the same $c_c\,c_d\,c_b^*$ and frequency.
Certain groups can drive a partner wave, $\kappa_X$, with the following frequency, dimensionless nonrelativistic energy, and magnetic quantum number,
\begin{equation} \label{eq:kinematics}
\omega_X = \omega_c + \omega_d - \omega_b, 
\qquad \epsilon \equiv e_c + e_d - e_b, 
\qquad m_X = m_c + m_d - m_b ,
\end{equation}
where $\omega_a \simeq \mu + E_a$ is the frequency of level $a$, rest mass included, and $\epsilon$ is the nonrelativistic energy, rest mass excluded, in units of $\mu \alpha^2$.
The partner wave is the generalization of $X$ that entered in the transition $211 + 211 \to 322 + X$ of the main text.

The groups with $b = c$ or $b = d$ oscillate at a frequency set by one of the cloud levels.
They act only to shift that level's frequency, which we neglect as in the main text.
We denote each remaining unique combination of coefficients and frequencies as $c + d \rightarrow b + X$, referring to it as self-scattering when $c = d$ and cross-scattering when $c \neq d$.
Writing the field as the cloud plus the partners and keeping the terms linear in the partners, each partner obeys
\begin{equation} \label{eq:partner}
\big(E_c + E_d - E_b - H_0\big)\,\kappa_X(\mathbf x) = c_c\,c_d\,c_b^*\,\frac{G\mu^3\alpha}{a_0^{3/2}}\,S(\mathbf x),
\qquad
S = \begin{cases}
F_c\,\hat\Phi_{bc}, & c = d,\\[2pt]
F_d\,\hat\Phi_{bc} + F_c\,\hat\Phi_{bd}, & c \neq d,
\end{cases}
\end{equation}
which defines the dimensionless SG source $S$ that plays an identical role to the source for the Auger calculations of the main text.
Note that a unique source is defined for each channel, although we leave that dependence implicit.

The behavior of the partner wave is determined by the sign of $\epsilon$.
For $\epsilon > 0$, it is unbound and escapes to infinity, representing an emission channel, which we treat in Sec.~\ref{sm:sec:emission}.
For $\epsilon < 0$, it cannot escape and is absorbed at the BH horizon, which we treat in Sec.~\ref{sm:sec:absorption}.
The two cases are the direct generalizations of $322 + 322 \to 211 + \infty$ and $211 + 211 \to 322 + X$ of the main text.
We consider up to relatively large $n$, and Pythagorean triples can allow for processes like $1544 + 2044 \rightarrow 1244 + X$ where $\epsilon = 0$ exactly.
We find very few such processes appearing within our level set, and that the levels these channels draw on are never populated simultaneously.
Hence the rate of processes with partner waves carrying $\epsilon = 0$ should be very small, and we omit them.
Moreover, the fine and hyperfine corrections to the spectrum discussed in Sec.~\ref{sm:sec:resonances} spoil the exact cancellation of the energies.
This does not turn these rates off, but it returns each such channel to the emission or absorption case, according to the sign of the shifted $\epsilon$.

The source $S$ can be evaluated multipole by multipole.
For this, we first decompose the angular structure of $\hat\Phi_{bc}$.
We begin by factoring out the angular structure of the beat density, $F_b^*F_c = \rho_{bc}(x)\,Y_b^*Y_c$, with $\rho_{bc} = R_{n_b\ell_b}R_{n_c\ell_c}$, which contains the multipoles $Y_{L,m_c-m_b}$ with $|m_c-m_b| \le L$, $|\ell_b - \ell_c| \le L \le \ell_b + \ell_c$ by the triangle rule, and $\ell_b + \ell_c + L$ even by parity.
The angular weights are determined by
\begin{equation}
D^{bc}_L = \int d\Omega\, Y_{L,m_c-m_b}^*\,Y_b^*\,Y_c ,
\end{equation}
so that $\hat\Phi_{bc} = \sum_L \hat V^{bc}_L(x)\,Y_{L,m_c-m_b}$.
Solving for the $L = 1$ contribution yields
\begin{equation}
\hat V^{bc}_1(x) = -\frac{4\pi D^{bc}_1}{3}\Big[x^{-2}\int_0^x dy\, y^3\rho_{bc}(y)
- x\int_0^x dy\,\rho_{bc}(y)\Big],
\end{equation}
which receives contributions from both the cloud's gravitational potential and the recoil.
For $L \neq 1$, the recoil makes no contribution, and Poisson's equation gives
\begin{equation}
\hat V^{bc}_L(x) = -\frac{4\pi D^{bc}_L}{2L+1}\Big[x^{-L-1}\int_0^x dy\, y^{L+2}\rho_{bc}(y)
+ x^L\int_x^\infty dy\, y^{1-L}\rho_{bc}(y)\Big],
\end{equation}
which includes the monopole $L = 0$ when $\ell_b = \ell_c$ and $m_b = m_c$.
Inserting this multipole decomposition into Eq.~\eqref{eq:partner}, the source can be rewritten as
\begin{equation} \label{eq:Smult}
S(\mathbf x) = \sum_L \hat V^{bc}_L(x)\,R_{n_d\ell_d}(x)\,Y_{L,m_c-m_b}(\hat{\mathbf x})\,
Y_{\ell_dm_d}(\hat{\mathbf x}) + \big[c \leftrightarrow d\big],
\end{equation}
where the exchanged term is present only for cross-scattering.

\subsection{The Partner-Wave Profile}

The result in Eq.~\eqref{eq:Smult} provides a recipe for determining the source for a given channel through which we can determine the partner-wave profile, $\kappa_X$, using Eq.~\eqref{eq:partner}.
A convenient intermediate step is to expand both expressions in partial waves of the partner.
Writing $S = \sum_\ell S_\ell(x)\,Y_{\ell m_X}$, the source decomposes as
\begin{equation}
S_\ell(x) = \sum_L \hat V^{bc}_L(x)\,R_{n_d\ell_d}(x)\int d\Omega\, Y_{\ell m_X}^*\,Y_{L,m_c-m_b}\,Y_{\ell_dm_d}
+ \big[c \leftrightarrow d\big].
\end{equation}
The angular integrals are nonzero only for $|m_X| \le \ell \le \ell_b + \ell_c + \ell_d$ with $\ell \equiv \ell_b + \ell_c + \ell_d \pmod 2$.
These are the same selection rules as for a quartic contact interaction studied in Ref.~\cite{Baryakhtar:2020gao}, since both are effective two-body interactions, one local and one long-range.

We organize the expansion of the partner wave into its partial waves as follows,
\begin{equation} \label{eq:kappa}
\kappa_X(\mathbf x) = c_c\,c_d\,c_b^*\,\frac{G\mu^2}{\alpha \,a_0^{3/2}}\sum_\ell \frac{u_\ell(x)}{x}\,Y_{\ell m_X}(\hat{\mathbf x}),
\end{equation}
which introduces a dimensionless reduced radial function $u_\ell(x)$, defined by
\begin{equation}
u_\ell(x) = \frac{\alpha\,a_0^{3/2}}{c_c\,c_d\,c_b^*\,G\mu^2}\;x\int d\Omega\, Y^*_{\ell m_X}(\hat{\mathbf x})\,\kappa_X(\mathbf x).
\end{equation}
From Eq.~\eqref{eq:partner}, the differential equation governing each partial wave is then
\begin{equation} \label{eq:radial}
\Big[\epsilon + \frac12\frac{d^2}{dx^2} + \frac1x - \frac{\ell(\ell+1)}{2x^2}\Big]u_\ell(x) = x\,S_\ell(x),
\end{equation}
with $u_\ell$ regular at the origin, decaying at large $x$ for $\epsilon < 0$ and outgoing for $\epsilon > 0$.
The solution can be written as the superposition of responses to a point source,
\begin{equation} \label{eq:green_def}
u_\ell(x) = \int_0^\infty dx'\, G^\ell_\epsilon(x, x')\, x' S_\ell(x'), \qquad
\Big[\epsilon + \frac12\frac{\partial^2}{\partial x^2} + \frac1x - \frac{\ell(\ell+1)}{2x^2}\Big] G^\ell_\epsilon(x, x') = \delta(x - x'),
\end{equation}
where the Green's function $G^\ell_\epsilon$ obeys the same boundary conditions as $u_\ell$.

The homogeneous form of Eq.~\eqref{eq:radial} is the radial equation of hydrogen, whose solutions are known in closed form.
For $\epsilon > 0$, they are the Coulomb functions, and for $\epsilon < 0$ they are the Whittaker functions.
Through the method of variation of parameters, we can construct the Green's function directly from the homogeneous solutions.
Away from $x'$, the Green's function solves the homogeneous equation, so it is built from a solution $u_{\rm reg}$ that is regular at the origin and a solution $u_{\rm out}$ with the required large-$x$ behavior,
\begin{equation}
G^\ell_\epsilon(x, x') = \frac{2}{\mathcal W}\,u_{\rm reg}(x_<)\,u_{\rm out}(x_>), \qquad
\mathcal W = u_{\rm reg}\,u_{\rm out}' - u_{\rm reg}'\,u_{\rm out},
\label{eq:green}
\end{equation}
with $x_< = \min(x, x')$ and $x_> = \max(x, x')$, where $\mathcal{W}$ is the Wronskian, which is independent of $x$ by Abel's formula.
This is the construction of Ref.~\cite{Takahashi:2024fyq} in Bohr units.
We construct it explicitly for emission channels, $\epsilon > 0$, in Sec.~\ref{sm:sec:emission}, and for absorption channels, $\epsilon < 0$, in Sec.~\ref{sm:sec:absorption}.

\subsection{Emission Channel Rates}
\label{sm:sec:emission}

For $\epsilon > 0$, with $k = \sqrt{2\epsilon}$, $\eta = -1/k$ and $\rho = kx$, the homogeneous form of Eq.~\eqref{eq:radial} is the Coulomb wave equation,
\begin{equation}
\frac{d^2u}{d\rho^2} + \Big[1 - \frac{2\eta}{\rho} - \frac{\ell(\ell+1)}{\rho^2}\Big]u = 0 ,
\end{equation}
whose solutions are spanned by the regular and irregular Coulomb functions~\cite{NIST:DLMF}, with $F_\ell(\eta, \rho) \propto \rho^{\ell+1}$ and $G_\ell(\eta, \rho) \propto \rho^{-\ell}$ near the origin.
Regularity therefore selects $u_{\rm reg} = F_\ell$, and the outgoing condition selects the outgoing Coulomb function $u_{\rm out} = H^+_\ell(\eta, \rho) = G_\ell + iF_\ell$.
For these choices, $\mathcal W = -k$.
Evaluating the partner wave of Eq.~\eqref{eq:green_def} far outside the cloud, we find
\begin{equation} \label{eq:Il}
u_\ell(x \rightarrow \infty) = -\frac2k\,I_\ell\,H^+_\ell(\eta, kx), \qquad
I_\ell = \int_0^\infty dx\, F_\ell(\eta, kx)\,x\,S_\ell(x) .
\end{equation}
Since $|H^+_\ell| \rightarrow 1$ at large $x$, the partner of Eq.~\eqref{eq:Il} is an outgoing wave of momentum $k/a_0$ and velocity $v = k\alpha$.

The number flux through a sphere of radius $r = a_0x$ follows from Eq.~\eqref{eq:kappa},
\begin{equation}
v\,r^2\int d\Omega\,|\kappa_X|^2  = N_cN_dN_b\,\frac{4}{k}\,G^2\mu^5\sum_\ell|I_\ell|^2 .
\end{equation}
Each escaping partner removes one quantum from each of $c$ and $d$ and adds one to $b$, so for cross-scattering this flux gives
\begin{equation} \label{eq:Ginf}
\dot N_b = -\dot N_c = -\dot N_d = \bar{\Gamma}_\infty\,N_cN_dN_b, \qquad
\bar{\Gamma}_\infty = \frac{4}{k}\,\alpha^2\mu q^2 \sum_\ell |I_\ell|^2,
\end{equation}
where as in the main text $q = \mu/M$.
For self-scattering, each escaping partner instead removes two quanta from $c$, so
\begin{equation}
\dot N_b = \bar{\Gamma}_\infty\,N_c^2N_b, \qquad \dot N_c = -2\bar{\Gamma}_\infty\,N_c^2 N_b.
\end{equation}
This classical calculation gives the stimulated part of the rate, and quantizing the gained level adds the spontaneous term, $N_b \rightarrow N_b + 1$, as described in Sec.~\ref{sm:sec:rates}.

\subsection{Absorption Channel Rates}
\label{sm:sec:absorption}

For $\epsilon < 0$, with $\beta = \sqrt{-2\epsilon}$, $\nu = 1/\beta$ and $z = 2\beta x$, the homogeneous form of Eq.~\eqref{eq:radial} is instead Whittaker's equation (the Coulomb wave equation at negative energy),
\begin{equation}
\frac{d^2u}{dz^2} + \Big[-\frac14 + \frac{\nu}{z} + \frac{1/4 - (\ell + 1/2)^2}{z^2}\Big]u = 0,
\end{equation}
whose solutions are the Whittaker functions of index $\ell + 1/2$~\cite{NIST:DLMF}.
These behave as $z^{\ell+1}$ or $z^{-\ell}$ near the origin and as $z^\nu e^{-z/2}$ or $z^{-\nu}e^{z/2}$ at large $z$.
In this case, regularity and decay at infinity select $u_{\rm reg} = M_{\nu,\ell+1/2}(z)$ and $u_{\rm out} = W_{\nu,\ell+1/2}(z)$.
Their Wronskian and the associated Green's function are
\begin{equation} \label{eq:green_neg}
\mathcal W = -\frac{2\beta\,\Gamma(2\ell+2)}{\Gamma(\ell+1-\nu)}, \qquad
G^\ell_\epsilon(x, x') = -\frac{\Gamma(\ell+1-\nu)}{\beta\,\Gamma(2\ell+2)}\,M_{\nu,\ell+1/2}(2\beta x_<)\,
W_{\nu,\ell+1/2}(2\beta x_>).
\end{equation}
Evaluating the partner wave of Eq.~\eqref{eq:green_def} near the BH, we find
\begin{equation} \label{eq:al}
u_\ell(x \rightarrow 0) \simeq a_\ell\,x^{\ell+1}, \qquad
a_\ell = -\frac{2\,(2\beta)^\ell\,\Gamma(\ell+1-\nu)}{\Gamma(2\ell+2)}\int_0^\infty dx\,
W_{\nu,\ell+1/2}(2\beta x)\,x\,S_\ell(x).
\end{equation}
The coefficient $a_\ell$ is the near-horizon amplitude of the forced partner, and it sets the rate at which the BH absorbs the partner.

A field that behaves as $\psi \simeq A\,x^\ell\,Y_{\ell m}/a_0^{3/2}$ near the BH and oscillates at frequency $\omega$ is absorbed by the BH at the Detweiler rate
\begin{equation} \begin{gathered} \label{eq:flux}
R_{\rm abs} = |A|^2\,4\tilde r_+\,K_\ell\,g_{\ell m}\, \frac{\omega - m\Omega_H}{\mu}\,\alpha^{4\ell+5}\mu, \\
K_\ell = 2^{2\ell-1}\Big(\frac{\ell!}{(2\ell)!}\Big)^2, \qquad
g_{\ell m} = \prod_{k=1}^{\ell}\Big[k^2(1-\chi^2) + \Big(m\chi - 2\tilde r_+\alpha\frac{\omega}{\mu}\Big)^2\Big],
\end{gathered} \end{equation}
where $g_{\ell m}$ is the horizon factor, $\tilde r_+ = 1 + (1-\chi^2)^{1/2}$ is the horizon radius in units of $GM$, and $\Omega_H = \chi/(2GM\tilde r_+)$ is the angular velocity of the horizon~\cite{Detweiler:1980uk, Baumann:2019eav}.
By Eqs.~\eqref{eq:kappa} and~\eqref{eq:al}, each partial wave of the partner takes this form near the BH, with
\begin{equation}
A = c_c c_d c_b^*\,\frac{G\mu^2}{\alpha}\,a_\ell, \qquad m = m_X, \qquad \omega = \omega_X.
\end{equation}
Each absorbed partner removes one quantum from each of $c$ and $d$ and adds one to $b$, so $N_c$ and $N_d$ fall, and $N_b$ rises, at the absorption rate summed over partial waves.
For cross-scattering, this gives
\begin{equation} \label{eq:GH}
\dot N_b = -\dot N_c = -\dot N_d = \bar{\Gamma}_H\,N_cN_dN_b, \qquad
\bar{\Gamma}_H = 4\tilde r_+\,\alpha^2 \mu q^2 \sum_\ell K_\ell\,g_{\ell m_X}\,\frac{\omega_X - m_X\Omega_H}{\mu}\,\alpha^{4\ell+3}\,|a_\ell|^2.
\end{equation}
For self-scattering, each absorbed partner instead removes two quanta from $c$, so
\begin{equation}
\dot N_b = \bar{\Gamma}_H\,N_c^2N_b, \qquad \dot N_c = -2\bar{\Gamma}_H\,N_c^2N_b .
\end{equation}
As for emission, quantizing the gained level adds the spontaneous term, $N_b \rightarrow N_b + 1$, as in Sec.~\ref{sm:sec:rates}.
Partial waves with $\ell \ge 1$ reach the horizon suppressed by $\alpha^{4\ell}$, but a partner with $m_X \neq 0$ has no $s$-wave, and they are then the leading term.

Since Eq.~\eqref{eq:flux} has the sign of $\omega - m\Omega_H$, a superradiant partner, $\omega_X < m_X\Omega_H$, has $\bar{\Gamma}_H < 0$, and the BH supplies the partner rather than absorbing it.
The channel then runs backwards, $b + X \rightarrow c + d$, and extracts energy and angular momentum from the BH, as in Eq.~\eqref{eq:bh}.
This can be understood as scattering-mediated SR.

\section{Resonances}
\label{sm:sec:resonances}

Through its overall scaling by $\Gamma(\ell + 1 - \nu)$, the Green's function for absorption in Eq.~\eqref{eq:green_neg} diverges when $\nu$ is an integer $N \ge \ell + 1$.
Since $\nu = 1/\sqrt{-2\epsilon}$, the divergence occurs at $\epsilon = e_N \equiv -1/(2N^2)$, where the partner wave has the energy of the bound level $(N, \ell)$.
This divergence is made manifest in the eigenfunction expansion of the Green's function defined by Eq.~\eqref{eq:green_def}, which takes the form
\begin{equation}
G^\ell_\epsilon(x, x') = \sum_{N \ge \ell+1}\frac{u_{N\ell}(x)\,u_{N\ell}(x')}{\epsilon - e_N}
+ \int_0^\infty d\epsilon'\,\frac{u_{\epsilon'\ell}(x)\,u_{\epsilon'\ell}(x')}{\epsilon - \epsilon'},
\label{eq:GreenFunctionExpansion}
\end{equation}
where $u_{N\ell} = x\,R_{N\ell}(x)$ are the bound levels and $u_{\epsilon'\ell} = (2/\pi k')^{1/2}\,F_\ell(-1/k', k'x)$, with $k' = \sqrt{2\epsilon'}$, are the unbound states, normalized to $\delta(\epsilon' - \epsilon'')$.
Near the BH, the bound levels behave as $u_{N\ell} \simeq a_{N\ell}\,x^{\ell+1}$, with
\begin{equation}
a_{N\ell} = \Big(\frac{2}{N}\Big)^{\ell+3/2}\left[\frac{(N+\ell)!}{2N\,(N-\ell-1)!}\right]^{1/2}\frac{1}{(2\ell+1)!}.
\end{equation}
The near-horizon amplitude of Eq.~\eqref{eq:al} is therefore
\begin{equation}
a_\ell(\epsilon) = a_\ell^{\rm cont}(\epsilon) + \sum_{N \ge \ell+1} \frac{a_{N\ell}\,M_N}{\epsilon - e_N}, \qquad
M_N = \int_0^\infty dx\, u_{N\ell}(x)\,x\,S_\ell(x),
\end{equation}
where $a_\ell^{\rm cont}$ is the contribution of the unbound states, which varies smoothly with $\epsilon$ for $\epsilon < 0$.
The residue of each pole is the product of the source's overlap with the level, $M_N$, which sets how strongly the channel drives it, and $a_{N\ell}$, which sets how much of the level reaches the BH.
Since the poles lie at $\epsilon < 0$, only absorption channels can be resonant.

When restricted to the two-level model, the partner wave in the $211 + 211 \rightarrow 322 + X$ channel has energy $\epsilon = -7/36$.
Then $\nu \simeq 1.6$, so the channel is nominally far from every pole.
However, in multi-level models, absorption channels are often at or near a pole.
In general, a channel $c + d \rightarrow b + X$ drives a partner with the energy $\epsilon$ and magnetic number $m_X$ of Eq.~\eqref{eq:kinematics}, in the partial waves $|m_X| \le \ell \le \ell_b + \ell_c + \ell_d$ with $\ell \equiv \ell_b + \ell_c + \ell_d \pmod 2$.
Since the hydrogenic spectrum depends on $n$ alone, the partner is exactly degenerate with a bound level whenever $b$ shares its principal number with $c$ or $d$, which gives $\epsilon = e_N$ with $N$ the principal number of the other.
For example, the partner of the self-scattering channel $944 + 944 \rightarrow 988 + X$ is exactly degenerate with the $900$, $920$, $940$, $960$, and $980$ levels, and that of the cross-scattering channel $211 + 644 \rightarrow 655 + X$ with $200$.
Other channels have accidental near degeneracies, such as $744 + 844 \rightarrow 1188 + X$, whose partner has $\nu \simeq 6.001$, near the $600$, $620$, and $640$ levels.
A channel at or near the pole of a bound level $(N, \ell, m_X)$ is a four-level process $c + d \rightarrow b + (N, \ell, m_X)$, followed by the decay of the intermediate level into the BH.

These divergences are regulated by the fine and hyperfine corrections to the Bohr spectrum~\cite{Baumann:2019eav} and by the width of the intermediate level.
Through $\mathcal{O}(\alpha^5)$, the frequency of a level $n\ell m$ is
\begin{equation}
\frac{\omega_{n\ell m}}{\mu} = 1 - \frac{\alpha^2}{2n^2} - \frac{\alpha^4}{8n^4} + \Big(\frac2n - \frac{6}{2\ell+1}\Big)\frac{\alpha^4}{n^3} + \frac{16}{2\ell(2\ell+1)(2\ell+2)}\,\chi m\,\frac{\alpha^5}{n^3}.
\end{equation}
The intermediate level decays into the BH at its Detweiler rate~\cite{Detweiler:1980uk, Baumann:2019eav} $\Gamma_N = -\Gamma_{N\ell m_X}$, so that its complex frequency is $\omega_N - i\Gamma_N/2$, with $\omega_N \equiv \omega_{N\ell m_X}$.
The relevant corrections to $\omega$ and $\Gamma$ are both higher order in $\alpha$ than the Bohr spacing, so they matter most near a pole, where they keep the rate finite.
As in the standard treatment of a resonance through a decaying intermediate state~\cite{Breit:1936zzb, Cohen-Tannoudji:1992}, we replace the hydrogenic detuning as follows
\begin{equation} \label{eq:prescription}
\epsilon - e_N \;\longrightarrow\; \Delta_N + \frac{i\,\Gamma_N}{2\mu\alpha^2}, \qquad
\Delta_N = \frac{\omega_c + \omega_d - \omega_b - \omega_N}{\mu\alpha^2},
\end{equation}
where all four frequencies include the fine and hyperfine corrections.
Each pole term is then the steady response of the intermediate level, driven at $\omega_c + \omega_d - \omega_b$ and damped by its decay into the BH.
The cloud's own gravitational potential also shifts each level, but again we have not included these shifts.
We first consider a channel near, but not at, a degeneracy.
We identify the bound level nearest the partner wave, and define the set $\mathcal{N}$ as that nearest bound level and the three levels on either side of it in energy.
Because the degeneracy is not exact, the unregulated $a_\ell$ calculated via Eq.~\eqref{eq:al} is finite, so it can be decomposed into the near-resonant contributors and a remainder as
\begin{equation}
a_\ell(\epsilon) = a_\ell^{\rm rem}(\epsilon) + \sum_{N \in \mathcal{N}} \frac{a_{N\ell}M_N}{\epsilon - e_N}.
\end{equation}
We regulate the poles of the near-resonant levels by the prescription of Eq.~\eqref{eq:prescription} to obtain the regulated amplitude
\begin{equation} \label{eq:atilde}
\tilde a_\ell = a_\ell^{\rm rem}(\epsilon) + \sum_{N \in \mathcal{N}} \frac{a_{N\ell}M_N}{\Delta_N + i\Gamma_N/2\mu\alpha^2}.
\end{equation}
Equivalently, the regulated amplitude is
\begin{equation} \label{eq:regulated_amplitude}
\tilde a_\ell = a_\ell(\epsilon) + \sum_{N \in \mathcal{N}} a_{N\ell}M_N \left[ \frac{1}{\Delta_N + i\Gamma_N/2\mu\alpha^2} - \frac{1}{\epsilon - e_N}\right].
\end{equation}
In principle, the pole regularization could be applied to more levels than the at most seven of $\mathcal{N}$, but at small $\alpha$ the energy differences across levels are large compared to the corrections of Eq.~\eqref{eq:prescription}.
As a result, we expect the contributions of additional levels to be negligible there.
The remainder and the residues $a_{N\ell}M_N$ depend only on the hydrogenic levels and are computed once per channel, while $\Delta_N$ and $\Gamma_N$ depend on $\alpha$ and $\chi$, so we evaluate Eq.~\eqref{eq:atilde} along each evolution.

An additional challenge arises in the case of an exact resonance, $\epsilon = e_{N_*}$ for some $N_* \in \mathcal{N}$, because Eq.~\eqref{eq:al} itself diverges there.
However, the remainder $a_\ell^{\rm rem}$ must be finite, since by construction it excludes the divergent resonant contributions, and it can be evaluated by
\begin{equation} \begin{split}
a_\ell^{\rm rem}(e_{N_*}) &= \lim_{\delta \to 0}\,\frac12\sum_{\pm}\Bigg[a_\ell(e_{N_*} \pm \delta)
- \sum_{N \in \mathcal{N}}\frac{a_{N\ell}M_N}{e_{N_*} \pm \delta - e_N}\Bigg] \\
&= \lim_{\delta \to 0}\,\frac12\sum_{\pm} a_\ell(e_{N_*} \pm \delta)
- \sum_{N \in \mathcal{N},\, N \neq N_*}\frac{a_{N\ell}M_N}{e_{N_*} - e_N},
\end{split} \end{equation}
where the $N_*$ pole term cancels between the two sides because it is odd in $\delta$.
The total regulated amplitude then follows straightforwardly from Eq.~\eqref{eq:atilde}.

In the full model we apply this regularization to every absorption channel so as to avoid ambiguity regarding when a near degeneracy is treated or not.
While these corrections are formally $\mathcal{O}(\alpha^2)$ or higher order, they can become $\mathcal{O}(1)$ corrections for $\alpha \gtrsim 0.2$ even for processes nominally far from resonance, indicating the necessity of a more detailed relativistic treatment.
Note also that the regularization prescription here assumes the steady response of a decaying intermediate level rather than a coherent passage through a level crossing, which would instead require a Landau--Zener treatment~\cite{Baumann:2019ztm}.
More generally, because we treat superradiance by only evolving occupation numbers of the levels without their associated phases, resonant coherent scattering is beyond the scope of our study.
We leave all three matters to future work.

\section{The two-level limit}
\label{sm:sec:twolevel}

We now consider the rates within the two-level model of the main text that restricted to only the $211$ and $322$ levels.
As mentioned in Sec.~\ref{sm:sec:setup}, the presentation of the calculation in the main text is distinct from that taken here.
In the main text the rates were computed as quantum-mechanical transitions in the inertial center-of-momentum frame, whereas here the computation is classical in the non-inertial frame of the BH.
The two approaches agree however, as this section demonstrates explicitly.

For the absorption scattering channel, $211 + 211 \rightarrow 322 + \mathrm{BH}$, we have $b = 322$ and $c = d = 211$.
The process is sourced by the beat density $F_b^*F_c = \rho_{bc}\,Y_{22}^*Y_{11}$, with radial part $\rho_{bc} = R_{21}R_{32} = x^3e^{-5x/6}/(243\sqrt5)$ and angular contributions only from the $L = 1$ and $3$ multipoles.
Equation~\eqref{eq:Smult} then gives
\begin{equation} \begin{gathered} \label{eq:S_abs}
S(\mathbf x) = R_{21}\,Y_{11}\,\big[\hat V^{bc}_1\,Y_{1,-1} + \hat V^{bc}_3\,Y_{3,-1}\big], \\
\hat V^{bc}_1(x) = \sqrt{\frac{8\pi}{15}}\,\Big[x^{-2}\!\int_0^x\! dy\,y^3\rho_{bc}(y) - x\!\int_0^x\! dy\,\rho_{bc}(y)\Big], \\
\hat V^{bc}_3(x) = -\sqrt{\frac{12\pi}{1715}}\,\Big[x^{-4}\!\int_0^x\! dy\,y^5\rho_{bc}(y) + x^3\!\int_x^\infty\! dy\,y^{-2}\rho_{bc}(y)\Big],
\end{gathered} \end{equation}
where $\hat V^{bc}_1$ contains the contribution from the recoil.
The partner wave has $\epsilon = -7/36$ and $m_X = 0$, so it is absorbed by the BH, at leading order in $\alpha$, through its $s$-wave.
Of the two products of harmonics in Eq.~\eqref{eq:S_abs}, only $Y_{11} Y_{1,-1}$ contains $Y_{00}$, with weight $-1/\sqrt{4\pi}$, so the $s$-wave source is $S_0 = -\hat V^{bc}_1R_{21}/\sqrt{4\pi}$.
Equation~\eqref{eq:al} with $\ell = 0$, $\beta = \sqrt{7/18}$, and $\nu = 1/\beta$ gives
\begin{equation}
a_{\ell=0} = -2\,\Gamma(1-\nu)\int_0^\infty dx\,W_{\nu,1/2}(2\beta x)\,x\,S_0(x) \simeq 0.1828,
\end{equation}
and Eq.~\eqref{eq:GH} with $K_0 = 1/2$, $g_{00} = 1$, and $\omega_X \simeq \mu$ then gives
\begin{equation}
\bar{\Gamma}_H = 2\tilde r_+ \mu\, \alpha^5 q^2 |a_{\ell=0}|^2 = \frac{A_H(\alpha\mu)^3}{\pi}\,q^2\,\Big|\frac{a_{\ell=0}}{2}\Big|^2,
\end{equation}
where the second form, with $A_H = 8\pi(GM)^2\tilde r_+$, matches Eq.~\eqref{eq:Gamma_H} at $\bar\lambda = 0$, with $\mathcal I_G = -a_{\ell=0}/2 \simeq -0.0914$.

For the emission scattering channel, $322 + 322 \rightarrow 211 + \infty$, we have $b = 211$ and $c = d = 322$.
The process is now sourced by the beat density $F_b^*F_c = \rho_{bc}\,Y_{11}^*Y_{22}$, although the radial part $\rho_{bc}$ is identical, and the angular part again only involves the $L = 1$ and $3$ multipoles.
From Eq.~\eqref{eq:Smult},
\begin{equation} \begin{gathered} \label{eq:S_em}
S(\mathbf x) = R_{32}\,Y_{22}\,\big[\hat V^{bc}_1\,Y_{11} + \hat V^{bc}_3\,Y_{31}\big], \\
\hat V^{bc}_1(x) = -\sqrt{\frac{8\pi}{15}}\,\Big[x^{-2}\!\int_0^x\! dy\,y^3\rho_{bc}(y) - x\!\int_0^x\! dy\,\rho_{bc}(y)\Big], \\
\hat V^{bc}_3(x) = \sqrt{\frac{12\pi}{1715}}\,\Big[x^{-4}\!\int_0^x\! dy\,y^5\rho_{bc}(y) + x^3\!\int_x^\infty\! dy\,y^{-2}\rho_{bc}(y)\Big].
\end{gathered} \end{equation}
The partner wave has $\epsilon = 1/72$ and $m_X = 3$, so it escapes to infinity.
Of the two products of harmonics in Eq.~\eqref{eq:S_em}, $Y_{11}Y_{22}$ contains only $Y_{33}$, while $Y_{31}Y_{22}$ contains $Y_{33}$ and $Y_{53}$, so the partner wave has two partial waves, with sources
\begin{equation}
S_3(x) = R_{32}(x)\Big[\frac{3}{2\sqrt{7\pi}}\,\hat V^{bc}_1(x) - \frac{1}{3\sqrt{2\pi}}\,\hat V^{bc}_3(x)\Big], \qquad
S_5(x) = \frac{5}{3\sqrt{22\pi}}\,R_{32}(x)\,\hat V^{bc}_3(x),
\end{equation}
of which only $\ell = 3$ receives the dipole and its recoil.
Equation~\eqref{eq:Il} with $k = \sqrt{2\epsilon} = 1/6$ and $\eta = -1/k = -6$ gives
\begin{equation}
I_\ell = \int_0^\infty dx\,F_\ell(-6, x/6)\,x\,S_\ell(x), \qquad I_3 \simeq 0.01532, \qquad I_5 \simeq 4.43\times10^{-4},
\end{equation}
and Eq.~\eqref{eq:Ginf} then gives
\begin{equation}
\bar{\Gamma}_\infty = 24\,\alpha^2 \mu\, q^2 \big(|I_3|^2 + |I_5|^2\big) \simeq 2\pi\alpha^2\mu\,q^2\,\Big|\sqrt{\tfrac{12}{\pi}}\,I_3\Big|^2.
\end{equation}
The second form drops the $0.08\%$ contribution from the $\ell=5$ partial wave.
It also exactly matches Eq.~\eqref{eq:Gamma_inf} at $\bar\lambda = 0$, with $\mathcal I^{(3)}_G = \sqrt{12/\pi}\,I_3 \simeq 0.02995$.

\section{Rate equations}
\label{sm:sec:rates}

Here we explain how the rate equations in Eqs.~\eqref{eq:rate_equations} and~\eqref{eq:MJ_evo} get modified for our multi-level treatment.
In short, we combine the SG rates as computed in Secs.~\ref{sm:sec:emission} and~\ref{sm:sec:absorption} with the SR rates, GW annihilations, and GW transitions into evolution equations for the occupation of the levels, as well as the mass and angular momentum of the BH.

Each SG channel $\sigma$, $c + d \rightarrow b + X$, proceeds at the rate
\begin{equation}
R_\sigma =
\begin{cases}
\bar{\Gamma}_\sigma\,N_c\,N_d\,(N_b + 1), & \bar{\Gamma}_\sigma \ge 0,\\[2pt]
-|\bar{\Gamma}_\sigma|\,N_b\,(N_c + 1)(N_d + 1), & \bar{\Gamma}_\sigma < 0,
\end{cases}
\end{equation}
with $\bar{\Gamma}_\sigma = \bar{\Gamma}_H$ or $\bar{\Gamma}_\infty$.
Each level grows or decays by SR at the Detweiler rate $\bar{\Gamma}_a$, with the full horizon factor of Eq.~\eqref{eq:flux} and the spontaneous term for a growing level,
\begin{equation}
\dot N^{\rm SR}_a = \begin{cases} \bar{\Gamma}_a\,(N_a + 1), & \bar{\Gamma}_a > 0,\\[2pt] \bar{\Gamma}_a\,N_a, & \bar{\Gamma}_a \le 0. \end{cases}
\end{equation}
Each level also annihilates to gravitons at the rate $\bar{\Gamma}_a^{\rm ann}$, and GW transitions $a \rightarrow a' + h$ proceed at $R_{a\to a'} =\bar{\Gamma}_{aa'}N_a(N_{a'} + 1)$, with the annihilation coefficients of Eq.~\eqref{eq:sr_rates} for $211$ and $322$, the flat-space annihilation coefficients of Ref.~\cite{Yoshino:2013ofa} for the other levels, and the flat-space transition rates of Refs.~\cite{Yoshino:2013ofa, Baryakhtar:2020gao}.
We neglect the cross-annihilation $a + a' \rightarrow h$ of different levels, which is suppressed relative to both the self-annihilation of each level and the transition between them, as in Ref.~\cite{Baryakhtar:2020gao}.
The occupations then evolve as
\begin{equation}
\dot N_a = \dot N^{\rm SR}_a + \sum_\sigma \nu_{a\sigma}\,R_\sigma - 2\bar{\Gamma}_a^{\rm ann}N_a^2 + \sum_{a'}\big(R_{a'\to a} - R_{a\to a'}\big),
\end{equation}
where $\nu_{a\sigma}$ is $+1$ for the gained level $b$, $-1$ for each of the lost levels $c$ and $d$, and $-2$ for $c$ in a self-scattering channel.

The BH evolves through SR and absorption of partner waves.
Each SR quantum extracts energy $\omega_a$ and angular momentum $m_a$, and each absorbed partner deposits $\omega_X$ and $m_X$, so
\begin{equation}
\dot M = -\sum_a \omega_a\,\dot N^{\rm SR}_a + \sum_{\sigma \in \rm abs}\omega_{X,\sigma}\,R_\sigma,
\qquad
\dot J = -\sum_a m_a\,\dot N^{\rm SR}_a + \sum_{\sigma \in \rm abs} m_{X,\sigma}\,R_\sigma,
\label{eq:bh}
\end{equation}
where the second sums run over the absorption channels.
In general, the absorbed partners deposit angular momentum of either sign, since $m_X = m_c + m_d - m_b$ is nonzero for most channels and takes both signs.

In the two-level model of the main text, with $211$ and $322$ denoted $1$ and $2$, the only scattering channels are those of Sec.~\ref{sm:sec:twolevel}, and the occupations obey
\begin{equation} \begin{aligned}
\dot N_1 &= \bar{\Gamma}_1 (N_1+1) - 2\bar{\Gamma}_H N_1^2 (N_2+1) + \bar{\Gamma}_\infty N_2^2 (N_1+1) - 2\bar{\Gamma}_1^{\rm ann} N_1^2, \\
\dot N_2 &= \bar{\Gamma}_2 (N_2+1) - 2\bar{\Gamma}_\infty N_2^2 (N_1+1) + \bar{\Gamma}_H N_1^2 (N_2+1) - 2\bar{\Gamma}_2^{\rm ann} N_2^2,
\end{aligned} \end{equation}
where we approximate $N_i(N_i-1) \simeq N_i^2$.
At leading order in $\alpha$, with $\omega \simeq \mu$, the SR and annihilation rates are
\begin{equation} \label{eq:sr_rates}
\bar{\Gamma}_{n\ell m} = 2\mu C_{n\ell}\,g_{\ell m}\,(m\chi - 2\alpha\tilde r_+)\,\alpha^{4\ell+4}, \qquad
\bar{\Gamma}^{\rm ann}_{n\ell} = \tfrac12\mu q D_{n\ell}\,\alpha^{4\ell+9},
\end{equation}
with $C_{21} = 1/48$ and $C_{32} = 4/885735$~\cite{Baumann:2019eav} and the $D_{n\ell}$ of the main text.
Since the only absorbed partner has $m_X = 0$ and $\omega_X \simeq \mu$, Eq.~\eqref{eq:bh} reduces to
\begin{equation}
\dot M = -\mu\big[\bar{\Gamma}_1 (N_1+1) + \bar{\Gamma}_2 (N_2+1)\big] + \mu\bar{\Gamma}_H N_1^2 (N_2+1), \qquad
\dot J = -\big[\bar{\Gamma}_1 (N_1+1) + 2\bar{\Gamma}_2 (N_2+1)\big],
\end{equation}
thereby fully reproducing the governing differential equations of the main text.

\section{Convergence testing}
\label{sm:sec:screen}

We test whether levels outside our model can affect the evolution by considering the addition of the levels $n\ell \ell$ with $\ell \le 16$ and $n\ell(\ell-1)$ with $\ell \le 8$, up to $n = 24$, and every level with $n \le 4$, corresponding to a total of $409$ new level candidates.
The 33-level model already includes the $n88$ levels because they can regulate the $n44$ levels and so backreact on $322$.
The candidates extend this to the further levels that the $n44$ and $n88$ levels populate, which could in turn regulate either.
Every level of the 33-level model has $m = \ell \le 8$, and a channel between two such levels, $n\ell \ell$ and $n'\ell'\ell'$, with a leading $s$-wave partner populates only the levels $NLL$ with $L = \ell + \ell' \le 16$.
Channels with higher partner waves are suppressed by further powers of $\alpha$, so the candidates with $\ell \le 16$ include every level up to $n = 24$ that the levels of our model populate.
We determine an upper bound for the occupation which might be realized in each of those candidate levels along the evolution of the BH in the 33-level model by evaluating
\begin{equation}
\dot N = A(t)\,N + B(t),
\end{equation}
where $A$ is the stimulated part, summing the candidate's signed SR rate, the SG channels by which the evolved levels of the 33-level model feed it, and the GW transitions into it from those levels, and $B$ is the spontaneous part of the same processes.
This is not a self-consistent evaluation of the candidate levels, but the bound is conservative because we neglect SG channels and GW transitions that would depopulate the candidate level.
The solution
\begin{equation}
N(t) = e^{\Lambda(t)}\Big[1 + \int_0^t dt'\,B(t')\,e^{-\Lambda(t')}\Big], \qquad
\Lambda(t) = \int_0^t dt'\,A(t'),
\end{equation}
then bounds the candidate's growth from a single quantum.
A candidate is added when its peak $\log N$ reaches a third of the $e$-folds it would need to matter, $\log(10^{-3}\,GM_0^2)$, about $57$ of $172$ for $M_0 = 10\,M_\odot$.
We then evolve the BH with the added levels and compare its spin history and $322$ occupation with those of the full model.

\begin{figure}[!t]
\includegraphics[width=\textwidth]{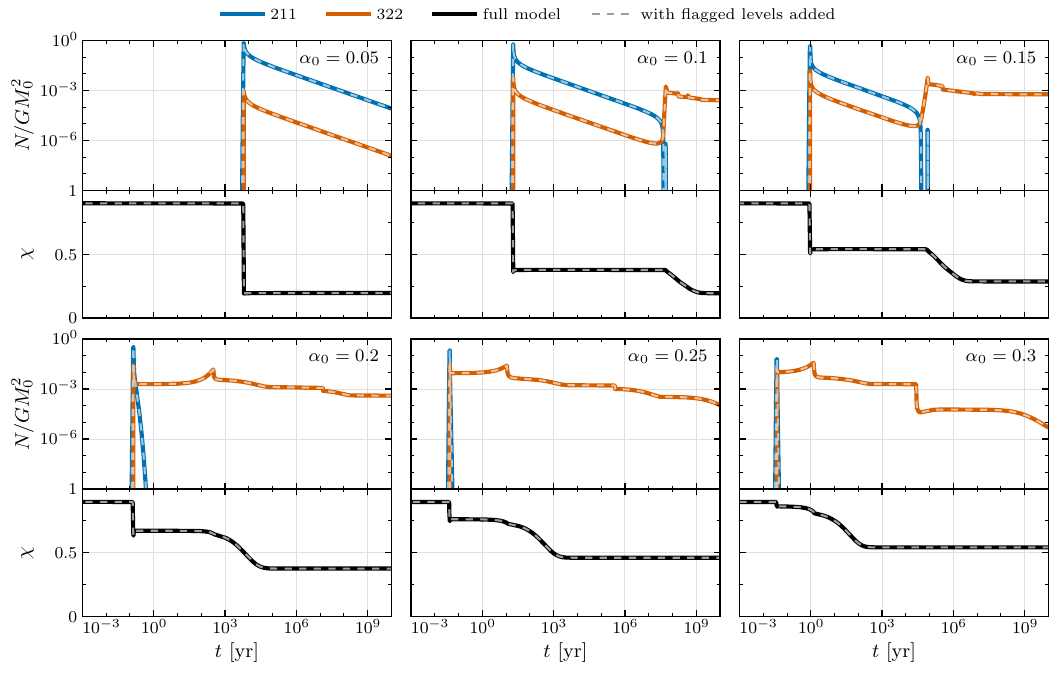}
\vspace{-0.8cm}
\caption{Evolution of the $211$ and $322$ occupations (upper panels) and the BH spin (lower panels) for BHs born with $M_0 = 10\,M_\odot$ and $\chi_0 = 0.9$ at six values of $\alpha_0$.
Solid curves show the full 33-level model and dashed curves the model extended by the levels flagged in the convergence test, which are indistinguishable.}
\label{fig:convergence}
\end{figure}

We perform this test on a grid of BHs with $M_0 = 10\,M_\odot$, pairing each of $\alpha_0 = 0.05$, $0.1$, $0.15$, $0.2$, $0.25$, and $0.3$ with each of $\chi_0 = 0.9$, $0.95$, and $0.99$.
We choose to focus on relatively large $\chi$ as being most relevant for generating observable $322$ lines.
For each pairing of $\alpha_0$ and $\chi_0$, we find excellent agreement, at the subpercent level at all times, in terms of the $211$ and $322$ occupations and the BH spin parameter $\chi$ between the 33-level full model and its augmented version with flagged levels added.
As an example, we illustrate the consistent results of the evolution with and without the flagged levels for $\chi_0 = 0.9$ in Fig.~\ref{fig:convergence}.
We choose to specifically focus on $\chi_0 = 0.9$, representing a direct convergence test of Fig.~\ref{fig:fullmodel} as presented in the main text, but similar levels of agreement are seen for $\chi_0 = 0.95$ and $\chi = 0.99$.

\end{document}